\documentclass[reprint,english,aps,prl,showpacs,superscriptaddress,preprintnumbers]{revtex4-1}

\usepackage[T1]{fontenc}
\usepackage[latin9]{inputenc}
\usepackage{float}
\usepackage{amsmath}
\usepackage{amssymb}
\usepackage{graphicx}
\usepackage{esint}
\usepackage{natbib}

\makeatletter
\@ifundefined{textcolor}{}
{%
 \definecolor{BLACK}{gray}{0}
 \definecolor{WHITE}{gray}{1}
 \definecolor{RED}{rgb}{1,0,0}
 \definecolor{GREEN}{rgb}{0,1,0}
 \definecolor{BLUE}{rgb}{0,0,1}
 \definecolor{CYAN}{cmyk}{1,0,0,0}
 \definecolor{MAGENTA}{cmyk}{0,1,0,0}
 \definecolor{YELLOW}{cmyk}{0,0,1,0}
}

\usepackage{amstext}
\usepackage{babel}

\addto\captionsenglish{}
\makeatother

\begin{document}
\title{A plasmonic nanoantenna based triggered single photon source}
\author{J. Straubel}
\affiliation{Institute of Theoretical Solid State Physics, Karlsruhe Institute of Technology, 76131 Karlsruhe, Germany}
\author{R. Filter}
\affiliation{Institute of Condensed Matter Theory and Solid State Optics, Abbe Center of Photonics, Friedrich-Schiller-University Jena, Max-Wien-Platz 1, 07743 Jena, Germany}
\author{C. Rockstuhl}
\affiliation{Institute of Theoretical Solid State Physics, Karlsruhe Institute of Technology, 76131 Karlsruhe, Germany}
\affiliation{Institute of Nanotechnology, Karlsruhe Institute of Technology, 76021 Karlsruhe, Germany}
\author{K. S\l owik}
\email{karolina@fizyka.umk.pl}
\affiliation{Institute of Theoretical Solid State Physics, Karlsruhe Institute of Technology, 76131 Karlsruhe, Germany}
\affiliation{Institute of Physics, Faculty of Physics, Astronomy and Informatics, Nicolaus Copernicus University, Grudziadzka 5, 87-100 Torun, Poland}

\begin{abstract}
Highly integrated single photon sources are key components in future quantum-optical circuits. 
Whereas the probabilistic generation of single photons can routinely be done by now, their triggered generation is a much greater challenge. 
Here, we describe the triggered generation of single photons in a hybrid plasmonic device. 
It consists of a lambda-type quantum emitter coupled to a multimode optical nanoantenna. 
For moderate interaction strengths between the subsystems, the description of the quantum optical evolution can be simplified by an adiabatic elimination of the electromagnetic fields of the nanoantenna modes. 
This leads to an insightful analysis of the emitter's dynamics, entails the opportunity to understand the physics of the device, and to identify parameter regimes for a desired operation. 
Even though the approach presented in this work is general, we consider a simple exemplary design of a plasmonic nanoantenna, made of two silver nanorods, suitable for triggered generation of single photons. 
The investigated device realizes single photons, triggered, potentially at high rates, and using low device volumes.
\end{abstract}


\maketitle
\section{Introduction\label{sec:introduction}}
The fabrication of nanoscopic systems has made great advancement during the last two decades. 
These advancements allowed to enter new regimes of light-matter-interaction, 
in which effects distinctive for cavity quantum electrodynamics could be realized at the nanoscale.
This aim can be achieved in the nearest future by exploiting on-chip integrated plasmonic nanoantennas, 
acting as open cavities coupled to adjacent quantum systems. 
The most prominent example of physical effects that have already been realized in this context
is the Purcell enhancement of the radiative decay rate of molecules or quantum dots adjacent to plasmonic nanoantennas (for a review please see Ref.~\onlinecite{Biagioni2012}).
The change in the transition rate due to a modified local density of electromagnetic states has been a subject of intensive studies on both 
experimental \cite{Anger2006,Tam2007,Curto2010,Bujak2011,Zhu2012,Akselrod2014} and theoretical grounds \cite{Dzsotjan2010,Schmidt2012, Filter2013}. 
Nanoantennas and nanowires that mediate the interaction between multiple emitters have also been discussed in numerous scenarios \cite{Dzsotjan2010,Slowik2013}, 
including entanglement generation protocols \cite{Martin-Cano2011,Chen2012,Hou2014}.

The key asset of nanoantennas is to tailor the light in the near- and far-field upon request. This includes shaping field distribution \cite{Li2007,Fischer2008, Filter2012}, directing light emission \cite{Staude2013,Krasnok2014,Coenen2014}, and tuning resonance frequencies \cite{Guo2008,Alu2008}
- all possibly with polarization-sensitive geometries. The scattered field can even possesses quantum properties, such as nonclassical statistics \cite{Benson2009,Claudon2010,Esteban2010,Maksymov2010,Busson2012,Filter2014,
MartinCano2015} or entanglement \cite{Nevet2010,Kivshar2012,Oka2013}, if suitable quantum emitters are exploited as sources. 
Such emitters, usually molecules and quantum dots, can be positioned at close vicinity of nanoantennas using the state-of-the-art techniques \cite{Bleuse2011,Chen_and_Chen2012,Chekini2015}.
Due to the broad character of the resonances in the optical domain, matching transition frequencies of the quantum emitters can be achieved with both dipolar, or even higher-order nanoantenna modes \cite{Alaee2015}.

\begin{figure}[t!]
\begin{centering}
\includegraphics[width=8.6cm,keepaspectratio]{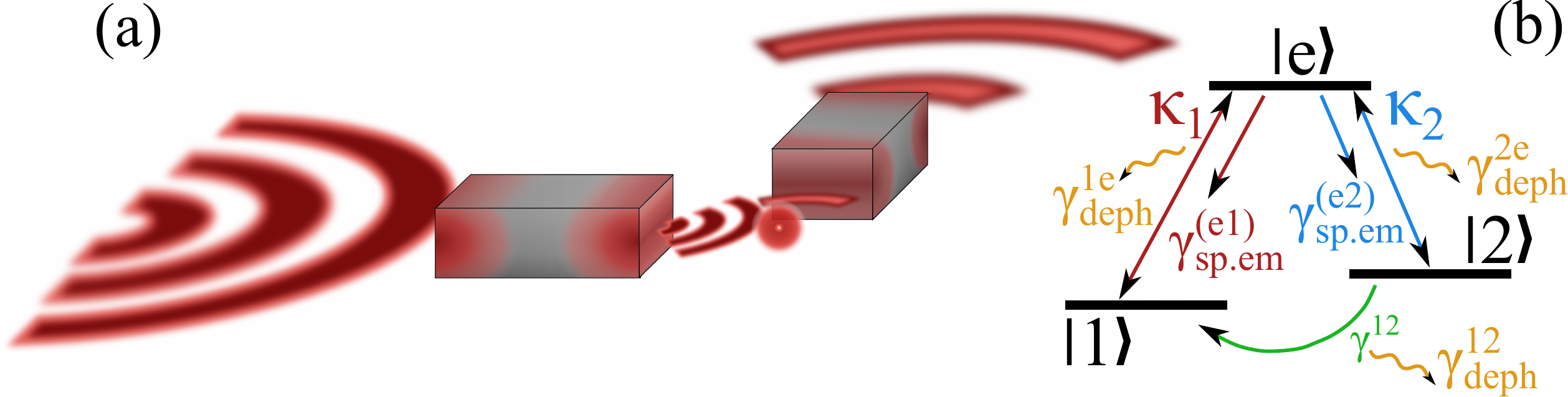}
\par
\end{centering}
\caption{\label{fig:setup_overview}(a) Scheme of the system under consideration: A two-modal nanoantenna coupled to a lambda-type quantum emitter.
(b) Jablonski diagram of the quantum emitter. 
Two-headed arrows indicate coherent, reversible transitions, while single-headed arrows correspond to incoherent, one-way transfer. 
Waivy arrows stand for dephasing, i.e. decay of the off-diagonal elements of the density matrix.}
\end{figure} 

In most of so-far proposed scenarios, quantum emitters were approximated as two-level systems. 
In such systems the transition probabilities to other energy eigenstates are assumed to be negligible. 
Considering richer energy configurations may, however, enable more complex dynamics. This allows promising effects such as nonlinear two-photon interactions \cite{Chang2007}.

However, in the context of interactions with quantum systems, 
there is still a distinct gap between what cavity quantum dynamics is able to predict as fundamental effects at the nanoscale and what can be achieved experimentally. 
With this paper our goal is to provide a stone to bridge this gap. 
We use a theoretical description of a carefully designed nanoantenna coupled to a realistic quantum system to yield new features that the are hardly observable with the isolated systems alone. 
We propose to exploit lambda-configuration quantum emitters coupled to bimodal nanoantennas for the conversion of light at the single-photon level. 
The unique advantage is the ability to achieve a triggered emission of a single-photon, i.e. the emission happens upon request \cite{Kuhn1999}. 
It will be shown that the extent of the deterministic character is limited by the nanoantenna efficiency; being however a parameter that can be engineered. 

A lambda-type emitter can represent a molecule or a quantum dot. Its configuration is given by two low- and one higher-energy state [Fig. \ref{fig:setup_overview}(b)]. Electric dipole transitions are assumed to be forbidden between the lower states, but allowed otherwise.

Suitable molecules have successfully been positioned in the vicinity of, or even adsorbed at a nanosurface, in numerous experiments in the context of surface-enhanced Raman scattering 
(for a review see Ref.~\onlinecite{Kumar2012}). 
Also, lambda-type and similar structures have recently been produced with self-assembled InAs quantum dots \cite{Dutt2005,Xu2008,Elzerman2011,Vora2015}. 
The exemplary nanoantenna design that we consider in this work is adjusted to support resonances in the frequency range and of polarization characteristics, corresponding to transitions in such quantum dots. 
Naturally, the nanoantenna can be redesigned for other target quantum systems.

In general, the nanoantenna does not necessarily require a specific geometry. On the contrary, various types of nanoantennas can be exploited, e.g. made from metallic or dielectric materials or even nanoantennas made from graphene \cite{Koppens2011,Grigorenko2012,Filter2013}. It is only important that the nanoantenna shall support multiple resonances, i.e. at least two. The resonances shall be sustained at the two transition frequencies of the lambda-type quantum emitter. For the simplified analytical description presented below, the nanoantenna should be weakly coupled to the quantum emitter. 
Then, multiple exchanges of energy among the subsystem are unlikely. This approximation is well justified for typical nanoantennas - it requires special effort to design a nanoantenna that operates beyond this limit \cite{Savasta2010,Slowik2013,Esteban2014}. 
Furthermore, we provide important figures of merit as definite analytical results for a simplified description of the effects of such hybrid quantum systems. 
Our approach to model the interaction of the nanoantenna and the lambda-type quantum system within the adiabatic limit is very general. However, we discuss an example of a metallic nanoantenna designed especially for the purpose of triggered single photon generation with lambda-type quantum emitters. 

Following discussions in the prior work \cite{Kuhn1999}, the ratio behind the design is the following. To achieve a triggered emission of a photon, a driving field is applied to the nanoantenna. The drive is resonant with the first transition of the lambda system. The drive excites the quantum emitter from its ground to its excited state. In the following, the emitter may relax to one of the two lower states - each of these processes is enhanced due to the coupling to a suitable resonance of the nanoantenna.
If the emitter returns to its ground state, the procedure quickly repeats due to the ongoing drive, i.e. typically at the nanosecond time scales. 
Otherwise, the desired single photon is generated in the transition to the other low-energy metastable state. 
This photon is then scattered into the far-field or absorbed by the nanoantenna at time scales of a few femtoseconds \cite{Maier2007}. 
Once the lambda system is in this second low-energy metastable state, the processes stop. We have witnessed the triggered emission of a single photon. 
The source might be continously used afterwards if a mechanism is applied that resets the quantum emitter into its ground state. 
This mechanism can be the slow internal decay from the metastable to the ground state, but it can also be enforced with, e.g., an optical pump. 
In this way the investigated device acts as an integrated source of triggered single photons, at rates controlable by the applied pump. 
The description of the functionality of such plasmonically enhanced triggered single photon source is at the heart of this contribution.

The paper is organized as follows. 
In Section \ref{sec:system} we introduce theoretical tools to explore the dynamics of the hybrid system of a quantum emitter 
and light modes supported by the nanoantenna. 
In Section \ref{sec:nanoantenna} we describe in detail a specific design of a nanoantenna that couples independently 
to the two transitions of the quantum emitter.
This simple design can be tuned in a straightforward manner to the target quantum system. 
Next, in Section \ref{sec:two_regimes}, a systematic analysis of the triggered emission of single photons is performed. 
The influence of numerous parameters of the hybrid system is investigated and clarified. 
In Section \ref{sec:drive}, we identify a transition between two distinguished regimes of the system's dynamics. 
While decreasing the driving field strength, the source loses its triggered character and the time of emission becomes random. 
The paper is summarized in Section \ref{sec:conclusions}. 
In the following Appendix A, we compare our proposal to a selection of other triggered single photon sources. 
In Appendix B, we develop a simplistic effective picture by adiabatically eliminating the electromagnetic fields. 
This provides an intuitive tool that hugely simplifies numerical simulations in the weak-coupling regime and grants a much better study of the physics. 
What is crucial, it allows to obtain the coupling constants between the fields and the quantum emitter. 
Finally, the adiabatic picture leads to necessary and sufficient conditions for an efficient triggered single-photon emission.

\section{Theoretical description of the hybrid system\label{sec:system}}
In this Section we introduce the system and discuss in detail the time-reversible processes described by its Hamiltonian, and various decay and decoherence mechanisms accounted for via the Lindblad formalism. 

The system consists of a quantum emitter in a lambda energy configuration [Fig.~\ref{fig:setup_overview}(b)]. 
The two low-energy states $|1\rangle$ and $|2\rangle$ are commonly referred to as "spin states". They may have different energies $\hbar \omega_{|1\rangle}$ and $\hbar \omega_{|2\rangle}$. We assume that state $|1\rangle$ is the ground state of the quantum emitter, in which the system is initially prepared. The excited state $|e\rangle$ has higher energy $\hbar \omega_{|\mathrm{e}\rangle}$. 
We additionally assume that the transitions between the states $|1\rangle$ ($|2\rangle$) and $|e\rangle$ are significantly faster 
than the direct transition between states $|1\rangle$ and $|2\rangle$. 
For instance, the first pair of transitions could be dipole-allowed, and the latter electric-dipole-forbidden.

The quantum emitter is coupled to a nanoantenna. Its optical response is characterized by scattering and absorption spectra. Here, we assume them to be well approximated by two Lorentzian resonances in the spectral region of interest. Since we will work at the single-photon level, we will apply the cavity-quantum-electrodynamics approach introduced in Ref.~\onlinecite{Waks2010} and represent the resonances as two quantum-mechanical bosonic modes, with annihilation operators $a_1$ and $a_2$. We assume that they are not directly coupled to each other, i.e. they are spectrally well separated or correspond to orthogonal polarizations. 
The modes are centered at frequencies $\omega_1$ and $\omega_2$, respectively. Due to frequency separation or due to polarizations, mode $\mathrm{j}$ ($\mathrm{j}=1,2$) couples only to the dipole-allowed transition between the state $|\mathrm{j}\rangle$ and the excited state $|\mathrm{e}\rangle$. The dipole-forbidden transition between the spin states is additionally assumed to be spectrally far-detuned from any resonance supported by the nanoantenna. The nanoantenna is illuminated by a laser of frequency $\omega_\mathrm{L}$, that directly drives mode $1$. The coupling to mode $2$ shall be negligible. All these requirements will be satisfied in excellent agreement by the explicit design we later suggest.

The corresponding Hamiltonian, in the frame rotating with the frequency of the driving field, reads:
\begin{eqnarray}
\mathcal{H}/\hbar &=& \left(\omega_{|\mathrm{e}\rangle}-\omega_\mathrm{L} \right)\sigma_{\mathrm{ee}}+\sum_{\mathrm{j}}\omega_{|{\mathrm{j}}\rangle}\sigma_{\mathrm{jj}}\\
&& +\sum_\mathrm{j}\left(\omega_{\mathrm{j}}-\omega_\mathrm{L}\right) a_{\mathrm{j}}^\dagger a_{\mathrm{j}}\nonumber \\
&& + \sum_{\mathrm{j}}\left(\kappa_{\mathrm{j}} a^\dagger_{\mathrm{j}}\sigma_{\mathrm{je}}+
\kappa_{\mathrm{j}}^\star \sigma_{\mathrm{ej}}a_{\mathrm{j}}\right) +\left(\Omega a_1^\dagger + \Omega^\star a_1\right), \nonumber
\end{eqnarray}
where $\sigma_{\mathrm{kl}} \equiv |{\mathrm{k}}\rangle\langle {\mathrm{l}}|$ is a flip operator between the quantum emitter's states. 
The first line in the above Hamiltonian corresponds to the free evolution of the quantum emitter. 
The second line represents the free dynamics of the field modes. 
The last line contains two terms: the first one stands for the coupling of the nanoantenna modes with the corresponding transitions in the quantum emitter. 
The coupling strength $\kappa_\mathrm{j}$ corresponds to the rate at which excitation is exchanged between the quantum emitter and mode $\mathrm{j}$ of the nanoantenna. 
A way to evaluate $\kappa_\mathrm{j}$ for a given nanoantenna design will be described in Section \ref{sec:nanoantenna}. 
The last term of the Hamiltonian represents the drive acting on mode $1$ of the nanoantenna with the strength $\Omega$, as described above. 
The value of $\Omega$ is proportional to the polarizability of the nanoantenna and to the electric field of the applied driving laser, as discussed in Ref.~\onlinecite{Slowik2013}.
Please note that we have neglected a direct coupling of the drive to the quantum emitter. 
This is an approximation rather than an additional requirement or complication to the scheme.
Due to generally large polarizabilities of nanoantennas, the plane wave drive couples to them significantly more efficiently than to quantum dots. Since the dot is position in the nanoantenna hot spot, the field which it experiences is dominated by the strong contribution scattered by the nanoantenna. 

The state of the full system (the quantum emitter and the two modes) is given by its time-dependent density matrix $\rho(t)$. The size of the density matrix is in principle infinite, but in our calculations we obtained stable results by restricting the Hilbert space to up to $10$ photons in mode $1$, and $5$ photons in mode $2$. The evolution of the system is governed by the Lindblad-Kossakowski equation \cite{Kossakowski1972,Lindblad1976}:
\begin{equation}\label{eq:lindblad_equation}
\dot{\rho}(t) = -\mathrm{i}/\hbar \left[\mathcal{H},\rho(t)\right]+\sum_\mathrm{p}\gamma_\mathrm{p}\mathcal{L}_{C_\mathrm{p}}\left[\rho(t)\right],
\end{equation}
where the Lindblad superoperators:
\begin{equation}\label{eq:linbdlad_operator}
\mathcal{L}_{C_\mathrm{p}}\left[\rho(t)\right] =
C_\mathrm{p}\rho(t)C_\mathrm{p}^\dagger - \frac{1}{2}\left(C_\mathrm{p}^\dagger C_\mathrm{p}\rho(t) + \rho(t)C_\mathrm{p}^\dagger C_\mathrm{p} \right)
\end{equation}
stands for various incoherence mechanisms in our system. They are represented by the corresponding operators $C_\mathrm{p}$ and rates $\gamma_\mathrm{p}$. We consider a number of processes that we will describe in the following. We will additionally provide orders of magnitude corresponding to the specific nanoantenna design that we discuss in Section \ref{sec:nanoantenna}, and for possible quantum systems. 

Usually, the fastest process in the hybrid system corresponds to losses by the nanoantenna. 
We consider radiative (scattering of photons into the far-field) and non-radiative contributions (absorption in the metal nanoantenna). 
Losses in mode $\mathrm{j}$ are given by the rate $\Gamma_\mathrm{j} = \Gamma^{(\mathrm{j})}_\mathrm{rad} + \Gamma^{(\mathrm{j})}_\mathrm{nonrad}$. 
For metallic nanoantennas, the typical order of magnitude is $10^{13}-10^{14}$ Hz (see Section \ref{sec:nanoantenna}, or Ref. \onlinecite{Chen2010}). 
For dielectric nanoantennas, the scattering rate is of the same order but it dominates over absorption \cite{Krasnok2012}. 
The scattering and absorption processes are described by Lindblad operators $\mathcal{L}_{a_\mathrm{j}}[\rho(t)]$. 

Spontaneous emission of the bare quantum emitter from the excited state to the state $|\mathrm{j}\rangle$ can be included via the term 
$\mathcal{L}_{\sigma_\mathrm{je}}[\rho(t)]$. 
In this way, the corresponding decoherence is naturally taken into account. 
The spontaneous emission rate is given by $\gamma_\mathrm{sp.em.}^{(\mathrm{ej})}$. 
Usually, this process is much slower than scattering and absorption in the nanoantenna modes: 
$\gamma_\mathrm{sp.em.}^{(\mathrm{ej})}\approx 10^7-10^9$ Hz \cite{Resch2008,Elzerman2011}. 
As we will show, the corresponding time-scale may be comparable to the time-scale at which nanoantenna-induced processes take place in the system. 
Therefore, this effect cannot be neglected. 

A similar process $\mathcal{L}_{\sigma_{12}}[\rho(t)]$ is related to nonradiative population transfer from the metastable state $|2\rangle$ 
to the ground state $|1\rangle$ (direct radiative transfer is dipole-forbidden). 
Normally, the corresponding intrinsic transfer rate $\gamma^{12}$ is an order of magnitude smaller than the above-described spontaneous emission rate \cite{Elzerman2011}. 
However, it can be boosted by pumping techniques.

The last process has to be taken into account if semiconductor quantum dots are considered as the quantum emitters. 
Pure decoherence or dephasing, i.e. decay of the coherence between states $|\mathrm{k}\rangle$ and $|\mathrm{l}\rangle$ in the quantum emitter, 
represented in an off-diagonal element of its density matrix, is given by the Lindblad term $\mathcal{L}_{\sigma_\mathrm{kk}-\sigma_\mathrm{ll}}[\rho(t)]$. 
In this process, the quantum-mechanical coherence is destroyed due to a coupling of the system to a fluctuating reservoir.
In general, the impact of this process grows with the size of the quantum system. 
In the solid which forms the quantum dot, the most important source of dephasing is usually the phononic bath: 
the system may acquire a random phase shift due to fluctuating lattice vibrations.
The dephasing rate between the excited and one of the low-energy levels $\gamma_\mathrm{deph}^\mathrm{je}$ varies for room temperatures between 
$10^9 - 10^{12}$ Hz \cite{Gammon1996,Bonadeo1996,Borri1999,Zrenner2000,Xu2008}. 
This is at least two orders of magnitude smaller than scattering and absorption in the nanoantenna, 
but can be comparable to coupling strengths $\kappa_\mathrm{j}$ and $\Omega$. 
The dephasing rate between the spin states is smaller $\gamma_\mathrm{deph}^{12}\approx 10^8$ Hz \cite{Xu2008}.
For a comprehensive discussion of quantum dephasing on an introductory or an advanced level, 
please see respectively the Refs.~\onlinecite{Marquardt2008} and \onlinecite{Shnirman2002}.

To summarize this section, we introduced here the theoretical tools within the Lindlad-Kossakowski formalism to describe the dynamics of 
a lambda-type quantum emitter in a vicinity of a nanoantenna that supports two orthogonal modes. 
In Appendix B, we perform an adiabatic elimination of the field that leads to a simplified picture valid in the regime of large nanoantenna scattering 
or absorption rates. This will be the regime in which the hybrid device will be operated.

In the following section, we propose a nanoantenna that supports modes in agreement with the previously formulated requirements. 
The nanoantenna will support two modes that are orthogonal in polarization and shifted in frequencies. 
In Section \ref{sec:two_regimes}, we will show that coupling a lambda-type quantum emitter to such nanoantenna will result 
in triggered single photon emission at nanosecond time-scales, if a sufficiently strong drive is applied. 

\section{Nanoantenna\label{sec:nanoantenna}}
In this section we describe a specific plasmonic nanoantenna that serves for the purpose of mode conversion and of the triggered generation of single photons. By simulating absorption and scattering spectra using a classical Maxwell solver, we calculate the related loss rates and the coupling constants to adjacent quantum emitters with predefined dipole moments. This methodology is standard and has been previously applied to describe single resonances \cite{Kivshar2012,Esteban2014}. 

\begin{figure}
\begin{centering}
\includegraphics[width=8.6cm,keepaspectratio]{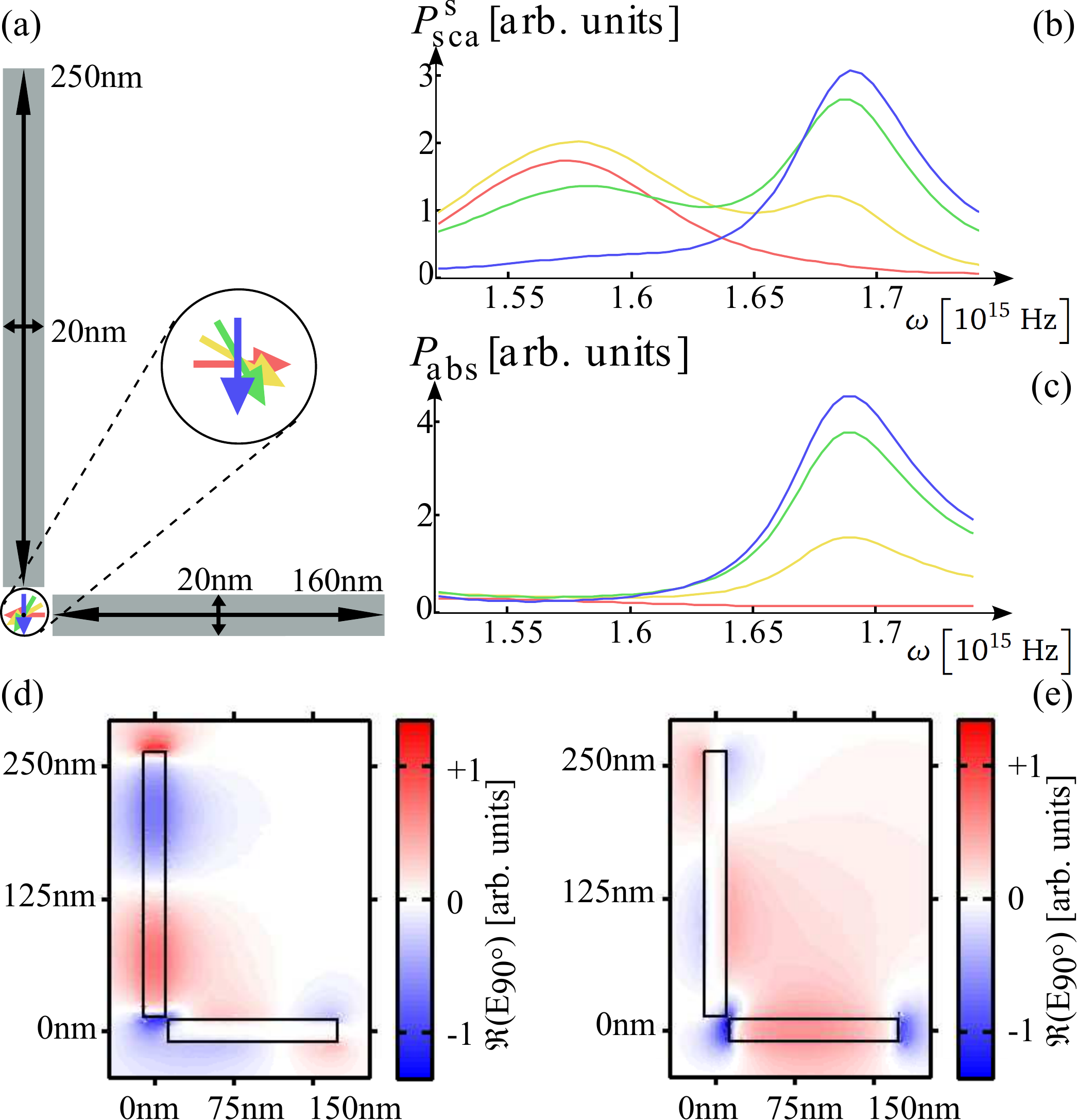}
\par
\end{centering}
\caption{\label{fig:nanoantenna}(a) Detailed scheme of the nanoantenna described in Section \ref{sec:nanoantenna}. The quantum emitter is placed at the position indicated by the arrows. Their orientations correspond to four orientations of the dipole moment that were used to simulate the scattering (b) and absorption (c) power spectra of the nanoantenna: blue line indicates a dipole emitter parallel to the longer nanorod, green, yellow, and red lines correspond to the dipole rotated by $30$, $60$ and $90$ degrees, respectively, as shown in panel (a). (d,e) Field distribution around the nanoantenna in the plane $5$ nm above the surface of the nanorods: (d) due to a dipole source at $\omega_1$ parallel to the longer nanorod; (e) due to a dipole source at $\omega_2$ parallel to the shorter nanorod. The component parrallel to the source orientation is shown.}
\end{figure} 

The proposed nanoantenna consists of two silver nanorods, perpendicular to each other [Fig.~\ref{fig:nanoantenna}(a)], embedded in glass with a permittivity of $\epsilon=2.25$. For simplicity, the rods are modelled as cuboids. 
Their geometrical parameters, i.e. lengths of $250$ and $160$ nm and heights and widths of $20$ nm each, have been adjusted to match with their resonances the infrared spectral region at which the transitions appear in the emitters described in Refs.~\onlinecite{Dutt2005,Xu2008,Elzerman2011,Vora2015}.
Naturally, the resonances can be tuned according to the spectral properties of the specific quantum emitters. The spectral position of the two modes and their overlap can be influenced by, e.g., a modification of the lengths of the two nanorods and their ratio. 
We are interested in the optical response of such nanoantenna subject to an excitation by a classical dipole positioned at a point equidistant from the tips of both nanorods, with a distance of $13.5$ nm. For a fixed frequency, such classical dipole approximates the behaviour of a corresponding transition of the quantum emitter. 

To simulate the scattering and absorption spectra of the nanoantenna, we solve the classical scattering problem in the frequency domain. The driving frequencies of the dipole source are fixed within a spectral range of interest. 
For this purpose, we have used the COMSOL Multiphysics simulation platform. The relative permittivity of silver based on the experimental data from Ref.~\onlinecite{Palik} has been considered. Postprocessing consists of calculating the absorbed $P_\mathrm{abs}\left(\omega\right)$ and scattered $P_\mathrm{sca}\left(\omega\right)$ powers. 
The former can be obtained by integration of the resistive losses:
\begin{equation}
P_\mathrm{abs}\left(\omega\right) = \int_V \bf{j}\left(\bf{r},\omega\right)\cdot\bf{E}\left(\bf{r},\omega\right) \mathit{dV},
\end{equation}
where the volume $V$ is restricted to the bulk of the nanoantenna. 
The scattered power is given by an integration of the outward normal Poynting vector over a closed surface $\bf{A}$:
\begin{equation}\label{eq:scattered_power}
P_\mathrm{sca}\left(\omega\right) = \int_{A} \mathbf{S}\left(\bf{r},\omega\right)\cdot d \bf{A}.
\end{equation}

In Fig.~\ref{fig:nanoantenna}(b), the simulated spectra $P_\mathrm{abs}\left(\omega\right)$ and $P^\mathrm{s}_\mathrm{sca}\left(\omega\right)$ are presented. Here, $P^\mathrm{s}_\mathrm{sca}\left(\omega\right)$ is the power related to the Poynting vector of the field scattered by the nanoantenna. 
The blue (red) line corresponds to the dipole moment of the emitter oriented horizontally (vertically). Green and yellow lines represent intermediate orientations. Clearly, the nanoantenna supports two orthogonally polarised modes. They are centered at $\omega_1 = 2 \pi\times 2.70 \times 10^{14}$ Hz and $\omega_2 = 2\pi\times 2.50 \times 10^{14}$ Hz, respectively. 
As follows from Fig.~\ref{fig:nanoantenna}(d) and (e), both modes differ in symmetry of their field distribution and we can consider mode $1$ as associated with the longer, and mode $2$ - with the shorter nanorod. This means that, to a good approximation, with the proper orientation of the dipole source we can address each mode individually. 
The perpendicular arrangement of the nanorods implies distinct radiation patterns into perpendicular directions for the two modes of interest. This entails the opportunity to distinguish the emitted photons by frequency, polarization, and propagation direction. 

To estimate the radiative and nonradiative loss rates by the nanoantenna, we fitted the scattering and absorption spectra, respectively, with Lorentzian line shapes. As a result, we obtained: $\Gamma^{(1)}_\mathrm{rad} = \Gamma^{(1)}_\mathrm{nonrad} = 6.8 \times 10^{13}$ Hz, $\Gamma^{(2)}_\mathrm{rad} = 1.0 \times 10^{14}$ Hz, $\Gamma^{(2)}_\mathrm{nonrad} = 2.2 \times 10^{14}$ Hz. Please note that the second mode is less pronounced in the absorption spectra since it is spectrally actually very broad. Its large width causes the second resonance to be rather flat, i.e. barely visible in the absorption spectrum. 

To estimate the coupling constant $\kappa_\mathrm{j}$, we exploit the resonant version of the adiabatic expression (\ref{eq:Purcell_rate}) from Appendix B:
\begin{equation} \label{eq:Purcell}
1+\eta^{(\mathrm{j})} \frac{4|\kappa_\mathrm{j}|^2}{\Gamma_\mathrm{j}\gamma^{(\mathrm{ej})}_\mathrm{sp.em}} = \frac{P^\mathrm{t}_\mathrm{sca}\left(\omega_\mathrm{j}\right)}{P^0_\mathrm{sca}\left(\omega_\mathrm{j}\right)}.
\end{equation}
On the right-hand side, we have expressed the same quantity as the ratio of the power related to the Poynting vector of the total field $P^\mathrm{t}_\mathrm{sca}\left(\omega_\mathrm{j}\right)$ and 
the one corresponding to the Poynting vector of the illumination field
$P^0_\mathrm{sca}\left(\omega_\mathrm{j}\right)$. Please note that Purcell enhancement \cite{Purcell1946, Haroche1989} is often understood as radiative enhancement of a localized point source into the far-field \cite{Tam2007}. In general this differs by the nanoantenna efficiency 
$\eta^\mathrm{(j)} \equiv \Gamma^\mathrm{(j)}_\mathrm{rad}/\left( \Gamma^\mathrm{(j)}_\mathrm{rad}+\Gamma^\mathrm{(j)}_\mathrm{nonrad}\right)$
from the actual decay rate enhancement of the quantum emitter, which in turn can be used to estimate the coupling constant.
For this reason we include the efficiency $\eta^{(\mathrm{j})}$ of the mode $\mathrm{j}$ into the expression (\ref{eq:Purcell}).
An efficiency deviating from unity accounts for the share of the energy extracted from the quantum emitter that is dissipated by the nanoantenna and will not be scattered into the far-field.
The spontaneous emission rate of a bare emitter can be expressed by the Weisskopf-Wigner formula \cite{Scully_Zubairy}
$$\gamma^{(\mathrm{ej})}_\mathrm{sp.em} = \frac{\left(\omega_{|e\rangle}-\omega_{|\mathrm{j}\rangle}\right)^3\sqrt{\epsilon}|d_\mathrm{ej}|^2}{3\pi\epsilon_0\hbar c^3}\ ,$$
with the vacuum speed of light $c$ and the transition dipole moment of $d_\mathrm{ej} = 6 \times 10^{-29}$ Cm. This leads to a rate consistent with experimental values \cite{Resch2008,Elzerman2011}.
With these assumptions, the coupling constants $\kappa_\mathrm{j}$ can be estimated to be almost equal, $\kappa_1 = 5.73 \times 10^{11}$ Hz and $\kappa_2 = 5.76 \times 10^{11}$ Hz respectively. Please note that such coupling constants are huge with respect to those typically obtained in conventional cavities, and therefore allow to address the quantum emitter with light extremely fast. Nevertheless, the ratio of the estimated coupling constants to the loss rates supports the approach that we have adopted here: the nanoantenna losses are indeed dominant, and therefore the adiabatic effective expression (\ref{eq:Purcell_rate}), that serves as a basis to estimate the coupling constants, is well justified and valid. 

With this exemplary design at hand, we can proceed to investigate the dynamics of the system modelled by the effective approach. 

\section{Triggered photon emission \label{sec:two_regimes}}
In this section, we will study the dynamics of the emission of a photon in mode $2$ for a lambda-type quantum emitter coupled to a nanoantenna described above.
We will show that for a driving field $\Omega$ of sufficient intensity and duration, a triggered single photon emission is achieved. 
First, we will investigate the dynamics of the single photon generation, where for simplicity we will start with a continous drive. 
Later, we emphasize the triggered character by studying the response to a pulsed excitation. 

\begin{figure}
\begin{centering}
\includegraphics[width=8.6cm,keepaspectratio]{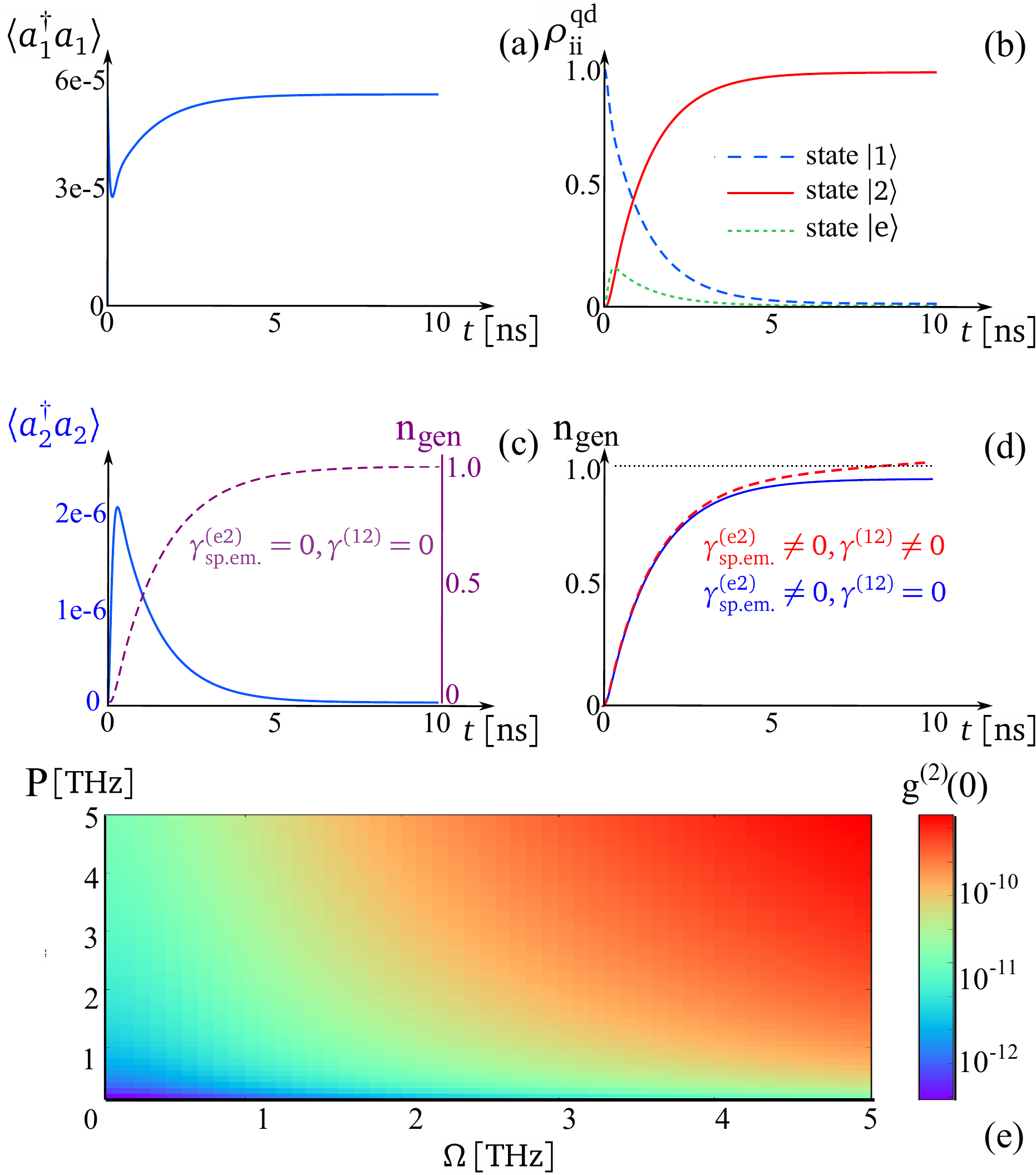}
\par
\end{centering}
\caption{\label{fig:dynamics}
Dynamics of a system consisting of a lambda-type quantum emitter coupled to a nanoantenna (described in Section \ref{sec:nanoantenna}): 
(a) mean photon number in mode $1$; 
(b) probability distribution of the quantum dot; 
(c) mean photon number in mode $2$ (blue solid line), 
and number of generated photons $n_\mathrm{gen}$ proportional to its integration (purple dashed line) 
for the case of neglected transfer rates $\gamma^{(\mathrm{e}2)}=\gamma^{(12)}=0$; 
(d) $n_\mathrm{gen}$ for included spontaneous emission rate $\gamma^{(\mathrm{e}2)}_\mathrm{sp.em.}=10^8$ Hz (blue solid line), 
and for additionally included transfer rate $\gamma^{(12)}=10^7$ Hz (red dashed line).
(e) Glauber second order correlation function at zero time delay $g^{(2)}(0)$ calculated for mode $2$ in function of the drive $\Omega$ and the pump $P$ 
proves the single-photon character of the generated field.}
\end{figure}

In Fig.~\ref{fig:dynamics}, we plot the dynamics of the emission process, given by the effective Lindblad equation (\ref{eq:lindblad_effective}), derived in Appendix A.
It is in good qualitative and quantitative agreement with the one obtained from the full Lindblad equation (\ref{eq:lindblad_equation}). 
In both cases, we have used the freely available QuTiP2 toolbox in Python \cite{qutip} for calculations. 
Mode $1$ of the nanoantenna is driven by a resonant laser field $\Omega = 5 \times 10^{11}$ Hz.
To assure the excitation of mode $1$ by the external drive, a possibility would be to use a plane wave polarized perpendicular to both nanorods. 
The direction of propagation should be at $92$ degrees relative to the short nanorod and at $-2$ degrees relative to the long nanorod. 
These slight deviations from an arrangement that adheres to the two nanorods are beneficial to suppress the excitation of mode $2$ by more than $10$ dB.
Moreover, the quantum emitter is suitably oriented such that the transition dipole moments of both transitions match the nanoantenna geometry. 
Both nanoantenna modes are assumed to be resonant with the quantum emitter's transition frequencies: 
$\omega_{|\mathrm{e}\rangle}-\omega_{|1\rangle} = \omega_1$, $\omega_{|\mathrm{e}\rangle}-\omega_{|2\rangle} = \omega_2$. 
For this simulation we take into account small, but realistic dephasing rates 
$\gamma_\mathrm{deph}^{1\mathrm{e}} =\gamma_\mathrm{deph}^{2\mathrm{e}} = 10\gamma_\mathrm{deph}^{12} = 10^9$ Hz 
\cite{Gammon1996,Bonadeo1996,Borri1999,Zrenner2000,Xu2008}.
We include spontaneous emission from the excited state to state $|1\rangle$ at rate: 
$\gamma_\mathrm{sp.em.}^{(\mathrm{e1})}=10^8$~Hz, but set the spontaneous emission to state $|2\rangle$ 
and the population transfer between the spin states $\gamma_\mathrm{sp.em.}^{(\mathrm{e2})},\gamma^{12}=0$. 
Their influence will be discussed later.

The quantum emitter is initially set in its ground state $|1\rangle$. 
Both nanoantenna modes are assumed to be in vacuum states. 
After switching on the drive, the mean number $\langle n_1 \rangle =\langle a_1^\dagger a_1 \rangle$  of photons in mode $1$ 
quickly reaches its stationary value [Fig. \ref{fig:dynamics}(a)], which is a result of a trade-off between driving and scattering and absorption. 
Note that the stationary value $\langle n_1 \rangle^\mathrm{steady} \lll 1$. 
This is due to huge losses characterizing metallic nanoantennas. 
However, this small value, that corresponds to a very rare presence of photons, 
is enough to drive the transition between states $|1\rangle$ and $|\mathrm{e}\rangle$ of the quantum emitter [Fig. \ref{fig:dynamics}(b)]. 
As the excited state gets populated, transfer to state $|2\rangle$ with a photon emission into mode $2$, 
becomes possible (here $\langle n_2 \rangle \equiv \langle a_2^\dagger a_2 \rangle$ 
is a very accurate approximation of the probability of a single photon in this mode). 
State $|2\rangle$ is the final state of the quantum emitter, 
as the mechanisms that could in principle drive a transition from state $|2\rangle$ to any other state, are very inefficient. 
The population transfer $\gamma^{12}$ due to internal collisions in the quantum emitter, has been neglected for this calculation. 
Due to huge values of scattering/absorption of the nanoantenna, the just-emitted photon quickly leaves the cavity, preventing re-absorption [Fig. \ref{fig:dynamics}(c)].

To confirm that such dynamics corresponds indeed to an emission of a single photon, we define the following function:
\begin{equation}
n_\mathrm{gen}\left( t \right)=\Gamma_2\int_0^t dt^\prime \langle a_2^\dagger a_2\rangle \left( t^\prime \right).
\end{equation}
For nanoantennas operating in the weak-coupling regime, this quantity can be interpreted as mean number of photons generated in mode $2$ by the time $t$. 
In the absence of transfer between the spin states $\gamma^{({12})}=0$, $n_\mathrm{gen}$ is equal to the probability of photon emission. 
In the idealized case of negligible free-space spontaneous emission to state $|2\rangle$, 
the quantum emitter can only make the transfer from $|e\rangle$ to $|2\rangle$ by emitting a photon in the desired mode $2$, 
and therefore $n_\mathrm{gen}\overset{t\rightarrow\infty}{\longrightarrow} 1$ [Fig.~\ref{fig:dynamics}(c), purple dashed line]. 
However, for $\gamma_\mathrm{sp.em.}^{(\mathrm{e2})}> 0$, the spontaneous emission channel opens, and the probability of photon emission drops. 
The blue solid line in Fig.~\ref{fig:dynamics}(d) corresponds to a realistic $\gamma_\mathrm{sp.em.}^{(\mathrm{e2})}=10^8$ Hz. 
This demonstrates that in case of metallic nanoantennas the correction to the probability due to spontaneous emission is rather small: 
the steady-state value of $n_\mathrm{gen}\left(t\rightarrow\infty\right) \approx 0.96$. 
(It might be of much greater importance for dielectric nanoantenas, whose Purcell enhancement is significantly smaller.) 
Finally, we include the process of internal population transfer between the spin states $\gamma^{(12)}=10^7$ Hz. 
Now it is possible that after the successful generation of a photon in mode $2$, 
the quantum emitter is reset to its ground state and the whole procedure repeats. 
As a result, the steady-state value of $\langle a_2^\dagger a_2\rangle \left(t\rightarrow\infty\right) >0$: 
it is about $1\%$ of its maximal value for the set of data analyzed here. 
Another photon may be generated in mode $2$, and therefore after the steady state is reached, 
$n_\mathrm{gen}$ grows linearly with a slope equal to $\langle a_2^\dagger a_2\rangle \left(t\rightarrow\infty\right)$ [Fig.~\ref{fig:dynamics}(d), red dashed line].
It is interesting to note that for our set of data the influences of  $\gamma_\mathrm{sp.em.}^{(\mathrm{e2})}$ and  $\gamma_\mathrm{sp.em.}^{({12})}$ 
happen to approximately cancel each other around the time when the system reaches its steady state. 

For completeness, we analyse the zero-delay second-order correlation function for mode $2$: 
$g^{(2)}(0) = \langle \left(a_2^\dagger\right)^2 \left(a_2\right)^2\rangle / \langle a_2^\dagger a_2 \rangle^2$. 
Its value is strictly equal to zero for single-photon Fock states, and can grow if a contribution from higher-number states is considerable.
For classical light $g^{(2)}(0)=1$. 
For the scenario proposed in this work the correlation function is equal to zero to a very good approximation, for 
physically realistic values of the drive $\Omega$ and the pump $P$ [Fig.~\ref{fig:dynamics}(e)].
First of all, this is because the final state of the dot is a metastable one, preventing two subsequent photon generation acts. 
In principle, a second act of photon generation might be enabled, if a fast pump $P$ from the final state $|2\rangle$ to the initial state $|1\rangle$
was applied to the lambda system, immediately recharging it after the first successful photon generation. 
However, even if such pump is applied, the second order correlation function will not grow, because of another reason:
the time necessary to produce a photon is much longer than its scattering/absorption time $1/\Gamma_2$, with $\Gamma_2\gg \kappa_\mathrm{j},\Omega$. 
In other words, the first photon is either emitted or absorbed before another one is produced. 

\begin{figure}
\begin{centering}
\includegraphics[width=8.6cm,keepaspectratio]{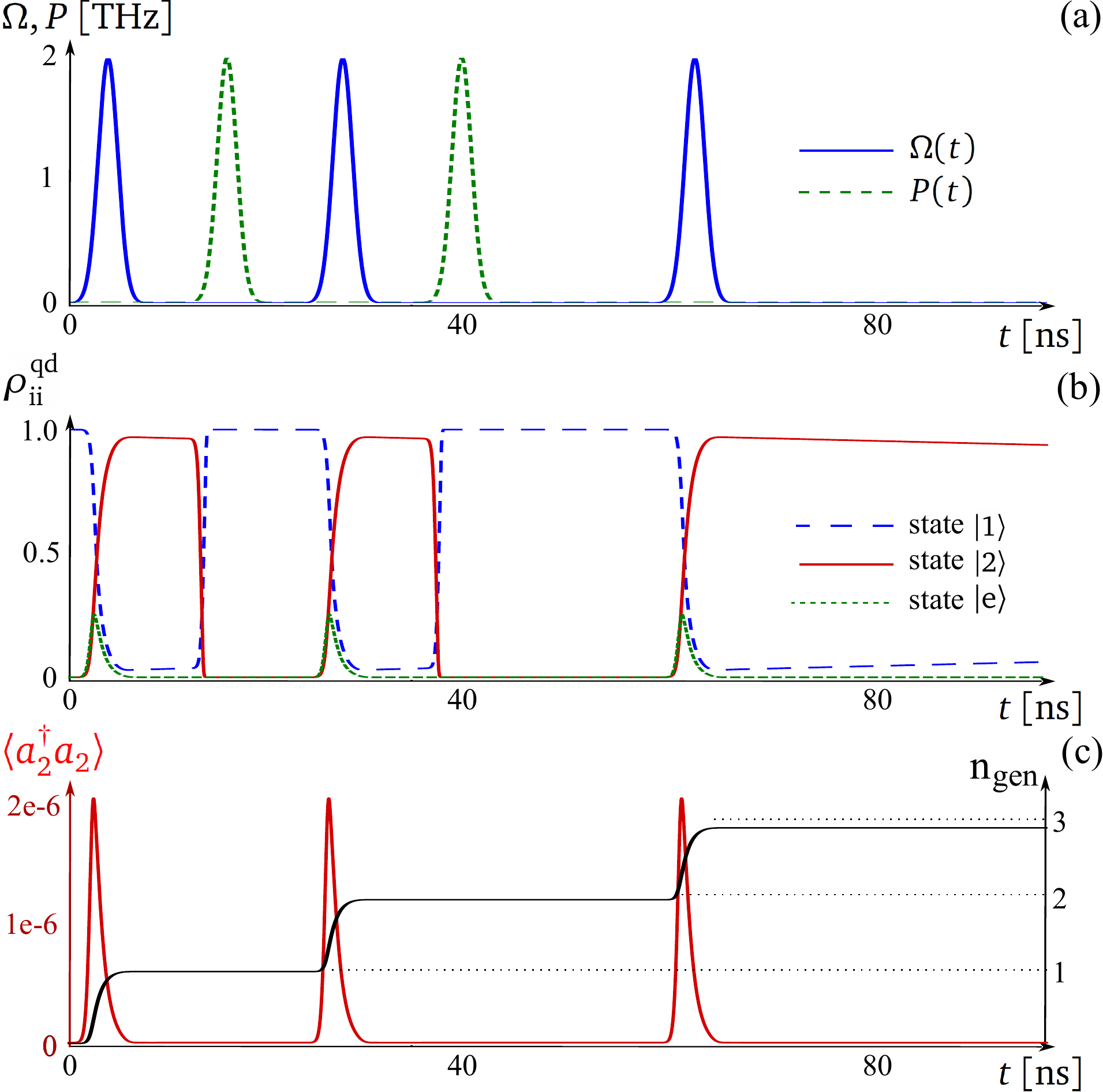}
\par
\end{centering}
\caption{\label{fig:pulses}A sequence of single photons emitted in the pulsed excitation scheme: (a) time-dependent drive $\Omega(t)$ and incoherent pump between the metastable states $P(t)$; (b) dynamics of the population distribution of the quantum emitter; (c) mean number $\langle a_2^\dagger a_2\rangle$ of photons in mode $2$ (red) and of generated photons $n_\mathrm{gen}$ (black line).}
\end{figure}

The photon generation in mode $2$ is thus complete. 
It is an on-demand process, i.e. it happens almost with certainty. 
However, it must be clearly stated that the photon may be afterwards emitted into the far-field with a probability equal to the nanoantenna efficiency $\eta^{(2)}$, 
or absorbed otherwise. 
The emission is triggered, i.e. it takes place in a short temporal window after the driving field $\Omega$ has been applied, i.e. at nanosecond time-scales.

To emphasize the triggered character of the emission, we analyze a pulsed excitation scheme, 
where a sequence of driving pulses of Gaussian temporal profiles is applied: 
$\Omega \left( t\right) = \Omega_0 \sum_i \exp\left[ -\left(t-t_i\right)^2/2\tau^2\right]$, 
of amplitude $\Omega_0=2$~THz, widths $\tau=1$~ns, centered at $t_1=4$~ns, $t_2=28$~ns, $t_3=64$~ns, respectively. 
In between, Gaussian pump pulses are applied that transfer the population from the metastable state $|2\rangle$ to the ground state 
$|1\rangle$: $P\left( t\right)= P_0\sum_i \exp\left[ -\left(t-\tilde{t}_i\right)^2/2\tau^2\right]$, 
of the same amplitude $P_0=2$~THz, same widths and centers at $\tilde{t}_1=16$~ns and $\tilde{t}_2=40$~ns [Fig.~\ref{fig:pulses}(a)]. 
We include the pump as an additional incoherent term $P(t)\mathcal{L}_{\sigma_{12}}\left[\rho^\mathrm{qd}(t)\right]$. 
All the other parameters are the same as above, and all decay and dephasing channels in the quantum emitter are included. 
Figure~\ref{fig:pulses}(b) illustrates how the population of the quantum emitter is flipped when the pulses are applied, 
leading with high probability to subsequent emissions of single photons directly after each application of a driving pulse [Fig.~\ref{fig:pulses}(c)]. 
The final value of $n_\mathrm{gen}\approx 2.94$, and would be equal to $3$ were the decay and dephasing in the quantum emitter neglected. 
This demonstrates that these processes play only a marginal role. 

In this section we have discussed in detail the performance of the proposed nanoantenna coupled to a lambda-type quantum emitter, as a triggered source of single photons. Below, we will show that the driving field $\Omega$ is the crucial parameter that controls the quality of the light source. 
Only for sufficient and reasonably large values, a triggered single photon emission is achieved. 
For too small values, the emission becomes random and looses its triggered character. 
The effective description introduced in Section \ref{sec:effective_description} will provide conditions that need to be fulfilled for an efficient, triggered emission of a single photon.
\section{Impact of the drive}\label{sec:drive}
\begin{figure}
\begin{centering}
\includegraphics[width=8.6cm,keepaspectratio]{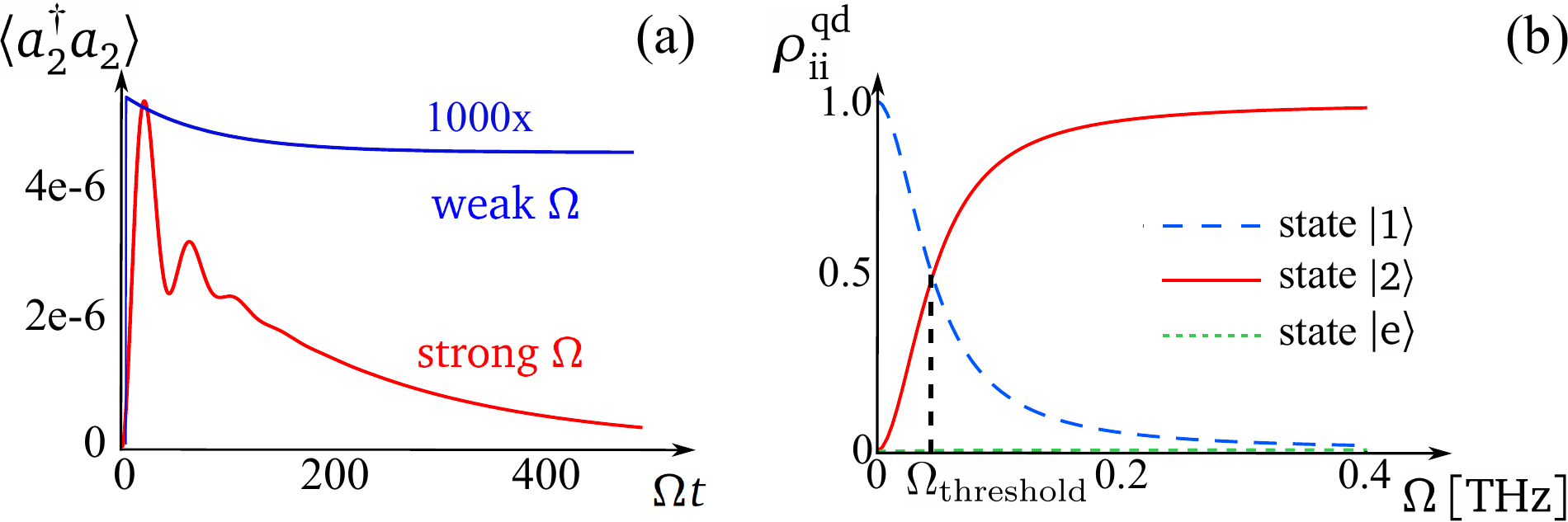}
\par
\end{centering}
\caption{\label{fig:dynamics_other_drives} (a) As in Fig. \ref{fig:dynamics}(c), but for a weaker $\Omega=2\times10^{10}$ Hz (blue line, multiplied $1000$ times) and stronger $\Omega=2\times10^{12}$ Hz (red line) drive. (b) Dependence of the stationary population distribution in the quantum emitter on the drive $\Omega$, for the nanoantenna parameters corresponding to the design of Section \ref{sec:nanoantenna}, and quantum-emitter parameters given at the beginning of Section \ref{sec:two_regimes}.}
\end{figure}

For the calculation in Fig.~\ref{fig:dynamics}, we have assumed a certain constant value of $\Omega = 5 \times 10^{11}$ Hz, 
that is strong enough for an efficient emission. 
As an indicator of successful single-photon generation we have taken the probability close to $1$ of the quantum dot being transferred to state $|2\rangle$. 
The efficiency of emission is, however, a complicated function of the parameters of the nanoantenna, location and orientation of the quantum dot, 
and of the driving field intensity. 
In a given experiment, the latter can still be tuned once the nanoantenna is produced and the quantum emitter positioned. 
To illustrate the impact, we have plotted in Fig.~\ref{fig:dynamics_other_drives}(a) 
the dynamics of the field in mode $2$ for a substantially weaker [$\Omega = 2\times 10^{10}$ Hz, blue line] 
and stronger [$\Omega = 2\times10^{12}$ Hz, red line] drive. 
In the former case, the excitation turns out to be too weak to successfully flip the quantum emitter into its state $|2\rangle$ 
and prevent its come-back to the initial state, leading to an emission of a series of photons at random times rather than a single triggered emission. 
We will refer to this case as the ``random-emission regime''.
In the latter case, the photon is efficiently generated, with a temporal shape that shows multiple peaks 
(their number increasing with the driving field), leading to a time-bin type emission \cite{Marcikic2002}. 

To identify a threshold the drive needs to overcome for efficient triggered photon generation, 
we have exploited the effective formalism described in the Appendix.
We have obtained a condition, which reads for the resonance $\delta_1=0$:
\begin{equation}\label{eq:condition}
|\Omega| \ggg \frac{\Gamma_1\sqrt{\Gamma_2 \gamma^{12}}}{2|\kappa_1\kappa_2|}\left(\frac{|\kappa_1|^2}{\Gamma_1}+\frac{|\kappa_2|^2}{\Gamma_2}\right),
\end{equation}
where we have assumed $|\kappa_\mathrm{j}|^2/\Gamma_\mathrm{j} \ggg \gamma^\mathrm{(ej)}_\mathrm{sp.em.}, \gamma_\mathrm{deph}^\mathrm{ij}$. 
For a sketch of the derivation, please see Appendix B.

The right-hand-side of the expression above is equal to approximately $\Omega_\mathrm{threshold} = 4\times 10^{10}$ Hz for the nanoantenna design discussed 
in Section \ref{sec:nanoantenna}. 
In Fig.~\ref{fig:dynamics_other_drives}(b), we plot the stationary population distribution of the quantum emitter as a function of the drive $\Omega$, 
for the nanoantenna parameters given in Section \ref{sec:nanoantenna}, and quantum emitter dephasing rates defined at the beginning of this section.  
The result was obtained in the effective formalism. 
Please note that the value of $\Omega_\mathrm{threshold}$ indicates approximately the driving field strength 
for which the stationary population is equally distributed between the spin states $|1\rangle$ and $|2\rangle$. 
However, to fulfill the condition (\ref{eq:condition}), the drive needs to be \textit{significantly stronger} than $\Omega_\mathrm{threshold}$.
This is in accordance with Fig.~\ref{fig:dynamics_other_drives}(b): 
the requirement of $\rho_{22}\approx 1$ is fulfilled for drives at least an order of magnitude larger.
Otherwise, if the driving field is too weak, a significant part of the stationary population remains in the ground state $|1\rangle$, 
from which it can be re-excited, leading to a sequence of photon emission acts in mode $2$ at random moments.
The device in such modus operandi might be considered for continuous conversion between nanoantenna modes via the quantum emitter. 

Please note that even though the condition in Eq.~(\ref{eq:condition_app}) has been derived in the effective formalism, 
the system does not fail to act as a single photon source for very strong drives, for which the effective formalism itself is no longer valid. 
This can be clearly seen from Fig.~\ref{fig:dynamics}(e) and its extensions to even larger $\Omega$s and $P$s, 
which has been obtained within the full Hamiltonian picture. 

In this section we have exploited the effective formalism to formulate an analytical condition of successful single-photon generation in the desired mode: 
switching between two working regimes of triggered and random-time photon emission may be possible by tuning the driving field intensity. 
For strong driving fields, a time-bin single photon can be produced.

\section{Conclusions\label{sec:conclusions}}
The scope of this work expands beyond the regime of single-mode nanoantennas, that is usually considered in the context of coupling to quantum emitters. Here, the multiresonant character of the nanoantenna spectrum is exploited for the purpose of inter-modal conversion. This is possible due to a coupling to a lambda-type quantum emitter. Specifically, we have considered the design of an L-shaped plasmonic nanoantenna, with a spectrum represented by two independent modes, centered at the transition frequencies of the quantum emitter. An analysis of these spectra has provided the parameters responsible for the dynamics of the coupled system. 

We have applied the effective description within the adiabatic approximation to find a quantitative distinction between two regimes of the dynamics, depending on the intensity of the drive. 
A driving field strong enough leads to a desired, triggered single-photon generation in a well-defined, short temporal window, directly after the application of the drive. 
Depending on the pump, such triggered emission may achieve high rates. 
A weaker drive results in a regime of continuous inter-modal conversion, with emission of subsequent quanta at random moments in time.

The problem of multimodal nanoantennas coupled to possibly multilevel quantum emitters should be further investigated in diverse contexts. Potential generalizations involve an analysis of light properties other than quantum statistics, e.g., entanglement in degrees of freedom related to particular modes. Furthermore, we have set the grounds for an important generalization to the case of a quantum-emitter transition coupled to more than one spectrally detuned nanoantenna mode. It will be the subject of our work in the nearest future. 

\section*{Appendix A: Parameter table and comparison\label{sec:table_compare}}
To give an overview over the most important parameters and most expressive properties of our proposed single photon generation scheme, 
we assembled these information and listed them in a tabular form. 
Furthermore in Table~\ref{tab:parametersANDcomparison} we juxtapose these data in opposition to 
both previously proposed single photon sources utilizing similar principles \cite{Kuhn1999} 
as well as state-of-the-art experimental single photon sources \cite{Ding2016, Rambach2016, Vora2015}.
Please note that Ref.~\onlinecite{Vora2015} is also based on lambda-type quantum dots. 

The comparison of the different parameters and properties displayed in Table~\ref{tab:parametersANDcomparison} 
clarifies both advantages and drawbacks of making use of a metal nanoantenna in combination with a lambda-type quantum emitter. 
First of all, the anticipated zero-delay second-order correlation function $g^{(2)}(0)$ is extremely small due to the weak-coupling regime our source operates at:
The cavity emission rate can be increased by multiple orders of magnitude in comparison to other single photon sources using the proposed setup. 
Additionally, the single photon rate as well as the cavity coupling constant exceed the performance data of alternative single photon generation schemes. 
The major disadvantage to all these benefits is the single photon extraction efficiency, 
which suffers naturally from the Ohmic losses of the metal nanoantenna. 
But even though these undesired, unavoidable losses are considered to be the dominating detriment 
to using metallic structures that support surface plasmon polaritons, 
the resulting extraction efficiency decreases only by a factor of 3 compared to the theoretical optimum without any Ohmic losses \cite{Kuhn1999}. 
This is in so far surprising as that one of the alternative experimental single photon generation setups possesses an even slightly lower efficiency, even though its operation is based on a very different physical mechanism. 

Finally, the proposed plasmonic single photon generation scheme provides further advantages. 
While the excitation of the lambda-type quantum emitter can be realized with classical resonant laser illumination, 
the trigger can be initiated by a resonant laser pulse, that is purely classical too. 
This allows the dynamic, external control of the single photons emission with smallest possible effort. 
Unlike some of the alternative single photon sources, the proposed scheme neither requires any $\pi$-pulse illumination nor any kind of cryostat. 
These should be obvious benefits regarding the experimental realization. 
And lastly the proposed setup is also at least one order of magnitude smaller in volume than any of the other single photon generation setups. 
This is why we consider it a very promising candidate for any kind of integrated photonic device relying on the triggered emission of single photons.

\begin{table*}[!tbh]
\caption{\label{tab:parametersANDcomparison} List of important parameters and expressive properties and comparison to other single photon sources.}
\begin{ruledtabular}
\begin{tabular}{l|rrrrr}
	parameter / & proposed design & Kuhn \textit{et al.} \cite{Kuhn1999} & Ding \textit{et al.} \cite{Ding2016} & Rambach \textit{et al.} \cite{Rambach2016} & Vora \textit{et al.} \cite{Vora2015}\\
	property    & theory          & theory                      & experiment                  & experiment                        & experiment\\
	\hline
	2nd order & & & & & \\
	correlation $g^{(2)}(0)$  & $\sim 10^{-11}$ & $-$ & $0.009$ & $-$ & $0.14$ \\
	\rule{0pt}{4ex}single photon  & & & & & \\
	emission rate  & $\left(\sim 3.3\times 10^{8} \mathrm{\ } \mathrm{Hz}\right)$\footnotemark[1] & $\sim 1.5\times 10^{6} \mathrm{\ } \mathrm{Hz}$ & $3.7 \times 10^{6} \mathrm{\ } \mathrm{Hz}$ & $1.2 \times 10^{8} \mathrm{\ } \mathrm{Hz}$ & $1.5 \times 10^{5} \mathrm{\ } \mathrm{Hz}$ \\
	\rule{0pt}{4ex}cavity emission & & & & & \\
	rate & $\sim 1.0 \times 10^{14} \mathrm{\ } \mathrm{Hz}$ & $\sim 2\pi \times 1.5 \times 10^{6} \mathrm{\ } \mathrm{Hz} $ & $-$ & $2.4 \times 10^{8} \mathrm{\ } \mathrm{Hz}$ & $4.6 \times 10^{10} \mathrm{\ } \mathrm{Hz}$ \\
	\rule{0pt}{4ex}cavity coupling  & & & & & \\
	constant  & $238.2 \mathrm{\ }\mu \mathrm{eV}$ & $0.013 \mathrm{\ }\mu \mathrm{eV}$ & $-$ & $-$ & $57 \mathrm{\ }\mu \mathrm{eV}$ \\
	\rule{0pt}{4ex}single photon & & & & & \\
	extraction efficiency & $\left(25.0\%\right)$\footnotemark[2] & $\left(73.6\%\right)$\footnotemark[2] & $66\%$ & $\left(23.2\%\right)$\footnotemark[2] & $-$ \\
	\rule{0pt}{4ex}external & classical excitation & pulsed excitation & & pumped by SHG- & \\
	excitation or & \& pulse to reset & \& pulse to reset & excitation & cavity driven by & pulsed laser \\
	control scheme & to initial state \footnotemark[3] & to initial state & by $\pi$-pulses & CW-laser & excitation \\
	\rule{0pt}{4ex}additional & variable trigger & & cryogen-free & multiple electro- & \\
	experimental & pulse length for & & bath cryostat & optic modulators & closed-cycle\\
	requirements & full functionality & $-$ & $\left(4.2\right.$ \ldots $\left.30\right)$ K & \& frequency locks & helium cryostat\\
	\rule{0pt}{4ex}approx. volume & L-shaped antenna, & high finesse cavity, & micropillar, & SPDC cavity, & photonic crystal, \\
	of setup & $\sim 0.003 \mathrm{\ } \mu\mathrm{m}^{3}$ & $\sim 0.01 \mathrm{\ } \mu\mathrm{m}^{3}$ & $\sim 177 \mathrm{\ } \mu\mathrm{m}^{3}$ & $> 10^{12} \mathrm{\ } \mu\mathrm{m}^{3}$ & $\sim 0.4 \mathrm{\ } \mu\mathrm{m}^{3}$ \\
\end{tabular}
\end{ruledtabular}
\footnotetext[1]{limited by spontaneous emission rate}
\footnotetext[2]{assuming a single photon detector with $80\%$ efficiency}
\footnotetext[3]{additional level of control: different behavior depending on choice of external drive and external trigger pulse}
\end{table*}

\section*{Appendix B: Effective hybrid system description\label{sec:effective_description}}
\setcounter{equation}{0}
\renewcommand{\theequation}{A\arabic{equation}}

The goal of this section is to derive an effective description of the investigated system within the adiabatic approximation. 
The reduced size of the effective Hilbert space hugely reduces the numerical simulation costs.
An even more important benefit is that this picture allows to obtain the coupling constants $\kappa_\mathrm{j}$. 
Moreover, it drastically simplifies analytical calculations, 
so that it becomes feasible to derive the condition (\ref{eq:condition}) on efficient single-photon generation. 

If a part of the physical system undergoes a decay at a rate much larger than other characteristic parameters of the dynamics, 
it can be effectively eliminated from the evolution \cite{Rice1988,Brion2007}. 
This is due to a mismatch in the time scales for the fast-decaying part and the remaining part of the system. 
With such elimination, the size of the physical problem can be drastically reduced. 
In the system investigated here, it is the electromagnetic field that undergoes huge scattering and absorption and which evolves at much shorter time scales. 
To obtain the effective description for the evolution of the quantum emitter, 
we integrate the equations of motion of the field operators in the Heisenberg picture and plug the resulting adiabatic expression back into the equations 
describing the evolution of the quantum emitter. 
In the last step, we transform the resulting equations back to the Schr\"odinger picture. 
A similar elimination has been explicitly described in Refs.~\onlinecite{Scully_Zubairy,Hou2014}. 

\begin{figure}[h]
\begin{centering}
\includegraphics[height=3cm,keepaspectratio]{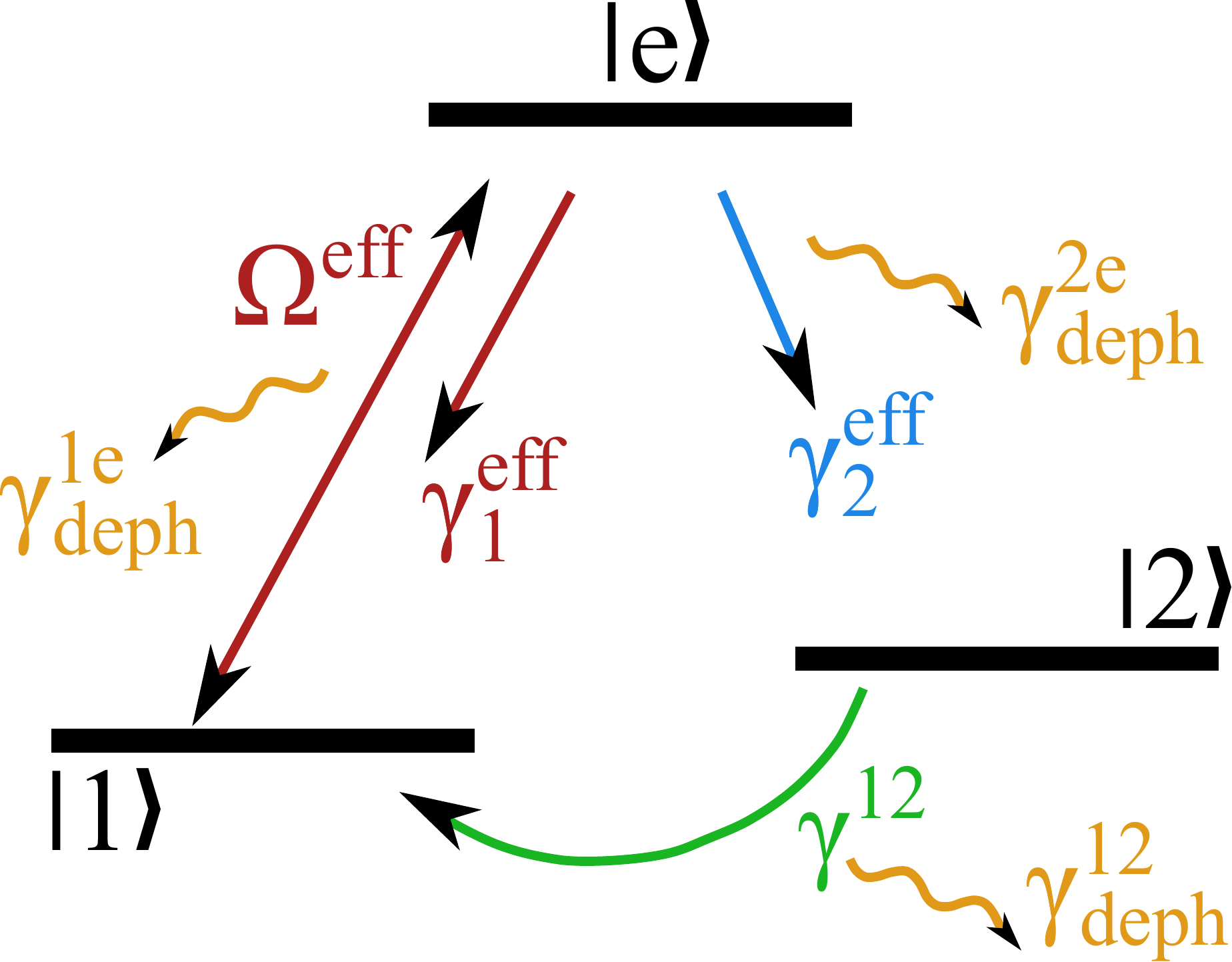}
\par
\end{centering}
\caption{\label{fig:appendix}The Jablonski diagram of the quantum emitter in the effective picture. Please compare to Fig.~\ref{fig:setup_overview}(b).}
\end{figure} 

Such procedure leads to a set of effective Bloch equations for the three-level quantum emitter solely. 
The effective Hamiltonian describes a coupling to a single coherent field driving the transition between the ground and the excited state [Fig.~\ref{fig:appendix}]:
\begin{eqnarray}
\mathcal{H}^\mathrm{eff}/\hbar &=& \left(\omega_{|\mathrm{e}\rangle}^\mathrm{eff}-\omega_\mathrm{L}\right)\sigma_\mathrm{ee}+\omega_{|1\rangle}\sigma_{11}+\omega_{|2\rangle}\sigma_{22} \\
&& + \Omega^\mathrm{eff}\sigma_{\mathrm{e}1}
+{\Omega^\mathrm{eff}}^\star\sigma_{1\mathrm{e}}, \nonumber
\end{eqnarray}
where 
\begin{equation}
\omega_{|\mathrm{e}\rangle}^\mathrm{eff} = \omega_{|\mathrm{e}\rangle} + \sum_{\mathrm{j}=1,2}\frac{|\kappa_\mathrm{j}|^2\left(\omega_{\mathrm{j}}-\omega_\mathrm{L}\right)}{\left(\omega_{\mathrm{j}}-\omega_\mathrm{L}\right)^2 + (\Gamma_\mathrm{j}/2)^2},
\end{equation}
is the energy of the excited state, shifted due to the nanoantenna field in a process analogous to the vacuum-induced Lamb-shift, and 
\begin{equation}
\Omega^\mathrm{eff} = \frac{-\mathrm{i}\kappa_1^\star\Omega}{\mathrm{i}\left(\omega_1-\omega_\mathrm{L}\right)+\Gamma_1/2},
\end{equation}
is the effective driving field.

Coupling to nanoantenna modes, originally coherent, leads in the effective picture to a modification of the population decay rates of the quantum emitter:
\begin{equation}\label{eq:Purcell_rate}
\gamma_\mathrm{j}^\mathrm{eff} = \gamma_\mathrm{sp.em.}^{(\mathrm{ej})} + \frac{|\kappa_\mathrm{j}|^2\Gamma_\mathrm{j}}{\left(\omega_{\mathrm{j}}-\omega_\mathrm{L}\right)^2 + (\Gamma_\mathrm{j}/2)^2}.
\end{equation} 
The process is effectively incoherent and accounted for in a Lindblad term, because the nanoantenna modes decay fast enough to prevent any back action, 
i.e. a reabsorption of the emitted photon.
At this point we would like to point out, that Purcell enhancement and modification of population decay rate concur in case of nanoantennas that do not suffer from Ohmic losses [see Eq. ~(\ref{eq:Purcell})].

The adiabatic dynamics can be found by solving the effective Lindblad-Kossakowski equation:
\begin{equation}\label{eq:lindblad_effective}
\dot{\rho}^\mathrm{qd}(t) = -\mathrm{i}/\hbar \left[\mathcal{H}^\mathrm{eff},\rho^\mathrm{qd}(t)\right]+\sum_\mathrm{p}\gamma_\mathrm{p}\mathcal{L}^\mathrm{eff}_{C_\mathrm{p}}\left[\rho^\mathrm{qd}(t)\right],
\end{equation}
where $\rho^\mathrm{qd}(t)$ is the density matrix of the quantum emitter. 
The effective Lindblad term includes the modified population decay rates $\gamma_\mathrm{j}^\mathrm{eff}$, 
as well as unchanged spin population decay $\gamma^{12}$, and dephasing rates $\gamma_\mathrm{deph}^\mathrm{ij}$.

\begin{figure}
\begin{centering}
\includegraphics[width=8.6cm,keepaspectratio]{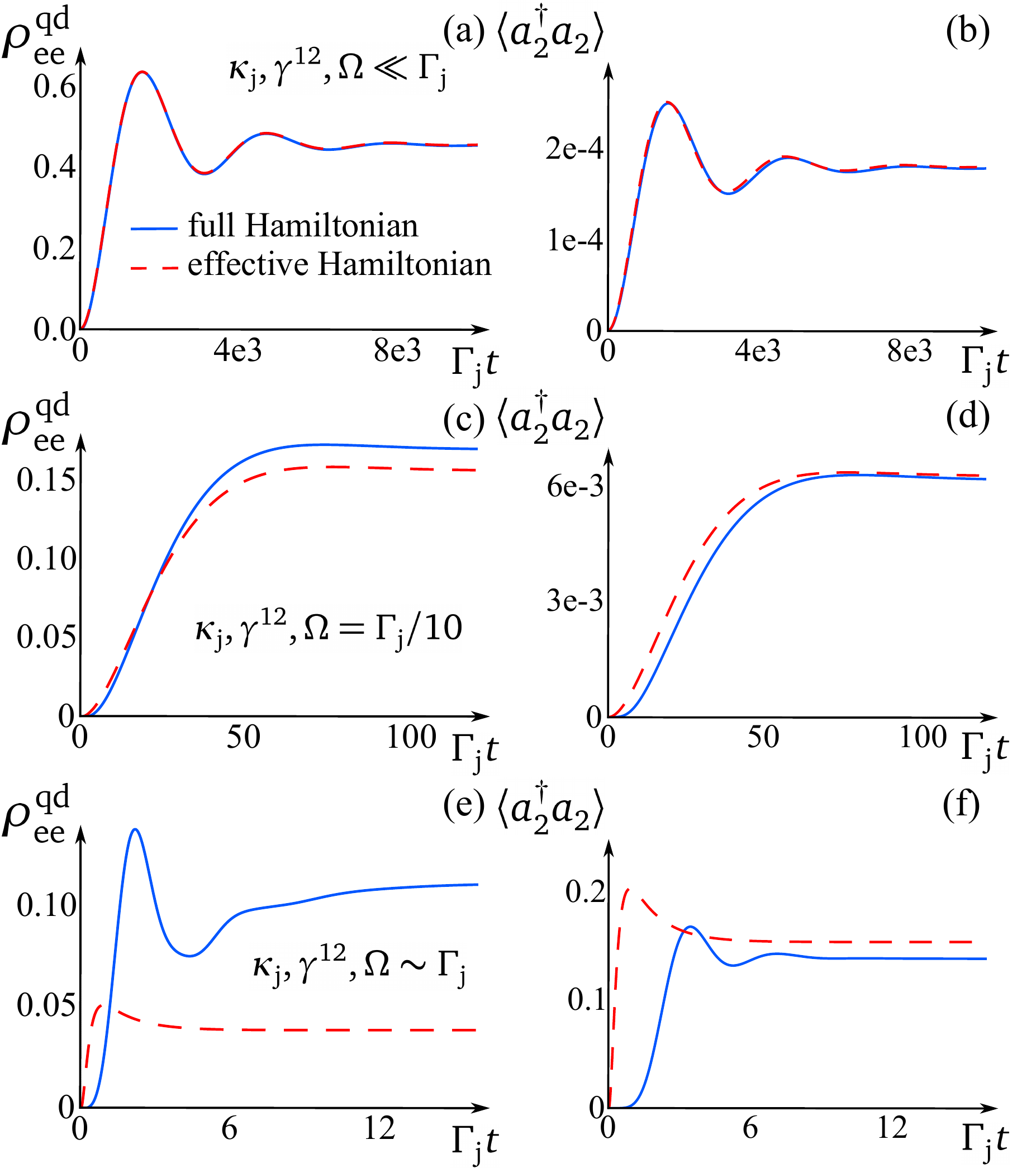}
\par
\end{centering}
\caption{\label{fig:validity}Evolution of the population in the state $|e\rangle$ (left) and photon number in mode $2$ (right) in the full-Hamiltonian 
and effective description. (a,b) Very good agreement for small $\gamma^{12},\Omega,\kappa_\mathrm{j}$, 
(c,d) qualitative, but not quantitative agreement in an intermediate case, 
(e,f) breakdown of the adiabatic approximation for values of parameters comparable to $\Gamma_\mathrm{j}$. 
Exact values of parameters chosen for simulations are provided in the main text.}
\end{figure} 

Additionally, we can track the excitation dynamics of the nanoantenna, 
following an adiabatic expression for the mean number of photons in each of the two modes:
\begin{eqnarray}
\langle a_1^\dagger a_1 \rangle(t) &=& \frac{|\kappa_1|^2\rho_{ee}(t)+\left[\Omega\kappa_1^\star\rho_{1e}(t)+\Omega^\star\kappa_1\rho_{e1}(t)\right]+|\Omega|^2}{\left(\omega_1-\omega_\mathrm{L}\right)^2+\left(\Gamma_1/2\right)^2},\nonumber \\
\langle a_2^\dagger a_2 \rangle 
(t) &=& \frac{ |\kappa_2|^2\rho_{ee}(t)}{\left(\omega_2-\omega_\mathrm{L}\right)^2+\left(\Gamma_2/2\right)^2}.
\end{eqnarray}

In Fig.~\ref{fig:validity}, we test the validity of the effective description, for simplicity assuming real $\kappa_\mathrm{j}$ and $\Omega$ parameters. 
The evolution of the excited state $|e\rangle$, as well as the photon number in the desired target mode $2$, 
calculated in the effective picture, are compared to those obtained from the full Lindblad-Kossakowski equation (\ref{eq:lindblad_equation}). 
For the validity analysis, we have chosen the population of state $|e\rangle$, 
because the difference between the results obtained in the two pictures is mostly manifested in this population. 
For this test, resonances in both transitions and of the driving field have been assumed, 
and decay and decoherence mechanisms of bare quantum emitters are neglected. 

We find almost perfect agreement as long as the coupling, pump, and drive is two orders of magnitude smaller than the nanoantenna loss rates, 
i.e. $\kappa_\mathrm{j} = \gamma^{12} = 0.01 \Gamma_\mathrm{j}$, $\Omega = 0.05\Gamma_\mathrm{j}$ [Fig. \ref{fig:validity}(a,b)]. 
A qualitative agreement, with slight differences in numbers, is obtained for parameters enhanced by an order of magnitude: 
$\kappa_\mathrm{j} = \gamma^{12} = \Omega = 0.1 \Gamma_\mathrm{j}$ [Fig. \ref{fig:validity}(c,d)]. 
If loss rates of the nanoantenna are comparable to other parameters $\kappa_\mathrm{j} = 2\gamma^{12} = 2\Omega = \Gamma_\mathrm{j}$, 
the validity of the adiabatic approximation breaks down [Fig. \ref{fig:validity}(e,f)] 
and predictions in the effective description do not properly predict the exact evolution. 

Similar conclusions have been reached while modifying other parameters, including dephasing rates, driving field, and detunings. 
In general, we have confirmed that the effective description accurately describes the dynamics when the scattering and absorption rates of the nanoantenna 
dominate among the parameters characterizing the system. 
This is in full agreement with the assumption of mismatched time-scales, made at the beginning of this section to justify the adiabatic elimination. 

We will now exploit the effective formalism to derive the threshold (\ref{eq:condition}), that the drive needs to overcome 
for efficient transfer of the quantum dot into the desired final state $|2\rangle$, which corresponds to a triggered photon generation.

From stationary effective Bloch equations, given by $\dot{\rho}^\mathrm{qd} = 0$, 
we derive the stationary population of state $|2\rangle$: $\rho^\mathrm{qd}_{22} = \left(1+\xi \right)^{-1}$, 
where $\rho^\mathrm{qd}$ stands for the stationary density matrix of the quantum dot in the effective picture. 
We have denoted
\begin{equation}
\xi \left(\Omega^\mathrm{eff}\right) = \frac{\gamma^{12}}{\gamma_2^\mathrm{eff}} 
\left[2+\frac{\left(\gamma_1^\mathrm{eff}+\gamma_2^\mathrm{eff}\right)\left(\Gamma_{1e}^2+4\delta_1^2\right)}{4\Gamma_{1e}|\Omega^\mathrm{eff}|^2} \right],
\end{equation}
with $\Gamma_{1e} = \gamma_1^\mathrm{eff}+\gamma_2^\mathrm{eff} + \gamma^{12}_\mathrm{deph}+4\gamma^{1e}_\mathrm{deph}
+\gamma^{2e}_\mathrm{deph}$, and $\delta_1 = \omega^\mathrm{eff}_{|e\rangle}-\omega_\mathrm{L}-\omega_{|1\rangle}$.
For an efficient transfer of the quantum dot into its state $|2\rangle$, we must have 
\begin{equation} \label{eq:condition_full}
\xi \left(\Omega^\mathrm{eff}\right)\lll 1.
\end{equation}
This can be understood as a condition that the driving field should fulfill. 
Typically, the ratio $\gamma^{12}/\gamma_2^\mathrm{eff} \approx 0.1$ in free space, and is even smaller in the presence of a nanoantenna. 
Therefore, condition (\ref{eq:condition_full}) can be regarded as always fulfilled if the driving field is strong $\Omega \rightarrow \infty$. 
For weak fields, we can simplify the above condition to
$|\Omega^\mathrm{eff}|^2\ggg \frac{\gamma^{12}}{4\gamma_2^\mathrm{eff}}\left(\gamma_1^\mathrm{eff}+
\gamma_2^\mathrm{eff}\right)\Gamma_{1e}$, where a resonance $\delta_1=0$ was assumed. 
In terms of original parameters, the condition of efficient state transfer can be approximately expressed in Eq.~(\ref{eq:condition}):
\begin{equation}\label{eq:condition_app}
|\Omega| \ggg \frac{\Gamma_1\sqrt{\Gamma_2 \gamma^{12}}}{2|\kappa_1\kappa_2|}\left(\frac{|\kappa_1|^2}{\Gamma_1}+\frac{|\kappa_2|^2}{\Gamma_2}\right),
\end{equation}
where we have considered the case of a nanoantenna-induced decay rate dominant over other decay and decoherence mechanisms in the quantum dot 
$|\kappa_\mathrm{j}|^2/\Gamma_\mathrm{j}\ggg \gamma^\mathrm{(ej)}_\mathrm{sp.em.}, \gamma_\mathrm{deph}^\mathrm{ij}$.

\section*{Acknowledgements}
The study was supported by a research fellowship within the project
``Enhancing Educational Potential of Nicolaus Copernicus University in the Disciplines of Mathematical and Natural Sciences'' (project no. POKL.04.01.01-00-081/10.) and the Karlsruhe School of Optics and Photonics (KSOP). Partial support by the German Science Foundation (within SPP 1391 Ultrafast Nano-optics) is acknowledged.

\bibliography{converter}
\end{document}